\documentstyle[twoside,fleqn,npb,epsfig]{article}

\newcommand{\AmS}{{\protect\the\textfont2
  A\kern-.1667em\lower.5ex\hbox{M}\kern-.125emS}}

\newcommand{\be}{\begin{equation}}
\newcommand{\ee}{\end{equation}}
\newcommand{\bea}{\begin{eqnarray}}
\newcommand{\eea}{\end{eqnarray}}

\newcommand{\non}{\nonumber}

\newcommand{\Fh}[2]{\,{}_#1F_#2}
\newcommand{\Fs}[3]{\!\!\left[\begin{array}{c}#1\,;\\#2\,;\end{array}#3\right]}

\newcommand{\Ffz}[2]{\Fs{#1}{#2}{z}}
\newcommand{\Ffe}[2]{\Fs{#1}{#2}{\frac{z}{z-1}}}

\newcommand{\Fde}[2]{\Fs{#1}{#2}{\frac{m_i^2}{r_{ij}}}}
\newcommand{\Fdz}[2]{\Fs{#1}{#2}{\frac{m_j^2}{r_{ij}}}}

\newcommand{\Fdi}[2]{\Fs{#1}{#2}{1-\frac{r_{ij}}{m_i^2} }}
\newcommand{\Fdj}[2]{\Fs{#1}{#2}{1-\frac{r_{ij}}{m_j^2} }}

\newcommand{\Fdf}[2]{\Fs{#1}{#2}{\frac{r_{ij}}{r_{ijk} }}}
\newcommand{\Fbf}[2]{\Fs{#1}{#2}{\frac{r_{ijk}}{r_{ijkl} }}}

\title{Application and explicit solution
 of  recurrence relations with respect to space-time
 dimension}

\author{O.V.~Tarasov\address{Fakult\"at f\"ur Physik,
Universit\"at Bielefeld,
D-33615, Bielefeld, Germany}%
        \thanks{On leave of absence from JINR,
Dubna (Moscow Region), Russian Federation.}
\thanks{
Supported by DFG}
\thanks{e-mail: tarasov@physik.uni-bielefeld.de}}

\begin{document}

\begin{abstract}
A short review of the method for the tensor reduction of Feynman
integrals based on recurrence relations w.r.t. space-time dimension
$d$  is given. A  solution of  the difference equation w.r.t. $d$ 
for the $n$-point one-loop integrals with arbitrary momenta and masses
is presented.  The result is written as multiple hypergeometric series
depending on  ratios of  Gram determinants. For the 3-point function
a new expression in terms of the  Appell hypergeometric function 
$F_1$ is presented.
\end{abstract}

\maketitle

\section{Introduction}
The increasing precision of current experiments and the
expected precision of future experiments reveal 
a need to  evaluate the two-loop and 
higher order radiative corrections \cite{erler}.
An evaluation of mass dependent two-loop corrections for many 
physical parameters such as the $\rho$ parameter
\cite{fjtro} and  $Z \rightarrow \overline{b}b$ decay width 
\cite{fjrt} is greatly desirable.
Of special interest is the problem of calculating  one-loop 
corrections to the process $e^+e^- \rightarrow 4$  fermions.
To solve this problem efficient algorithms
for evaluating multi leg one-loop tensor integrals are needed.
Some progress was recently achieved in this direction in
\cite{BDK}-- \cite{BGH}.

In the present paper we consider the algorithms based on recurrence
relations w.r.t.  space-time dimension or $d$-shift recurrence relations
which we hope will be useful for evaluating  multi loop
and multi leg integrals. In the first part we describe the
method for the reduction of tensor integrals to a set of basic 
integrals proposed in \cite{ovt1}.
In the second part we present a solution of recurrence relations
w.r.t.  $d$ for one-loop $n$-point scalar integrals 
with arbitrary  momenta and masses. This is a short report
on work done in collaboration with J.Fleischer and
F.Jegerlehner \cite{fjt00}.

\section{Reduction of tensor integrals}
The reduction of one-loop   tensor integrals to scalar integrals 
with shifted dimensions was given in \cite{Andrey}.
The reduction of multi loop tensor integrals was considered in \cite{ovt1}.
In the approach of \cite{ovt1} any tensor $L$ loop integral is 
represented as
\begin{eqnarray}
\prod_{i=1}^{L} \!\int \!\frac{d^dk_i}
{\pi^{dL/2}} \!\prod^{N}_{j=1}\!
P^{\nu_j}_{\overline{k}_j,m_j}\!\prod_{r=1}^{n_1}
\!k_{1\mu_r} \ldots \prod_{s=1}^{n_L}
k_{L\lambda_s}\! =\! \nonumber \\
T_{\mu_1 \ldots \lambda_{n_L}}\!(\{p_i\},\{ \partial_j\}, {\bf d^+})\!
\prod_{i=1}^{L}\! \int \frac{d^dk_i}{\pi^{dL/2}} \prod^{N}_{j=1}\!
P^{\nu_j}_{\overline{k}_j,m_j},
\label{tensint} 
\end{eqnarray}
where $\overline{k}_j$ is the momentum of the $j$-th line,
$p_i$ are external momenta,
$T_{\mu_1 \ldots \lambda_{n_L}}$ is a tensor operator and
\begin{eqnarray}
&&\partial_j \equiv \frac{\partial}{\partial m_j^2},
~~~~{\bf d^+} G^{(d)}=G^{(d+2)}, \nonumber \\
&&P_{k,m} = \frac{1}{k^2-m^2+i\epsilon}. \nonumber 
\end{eqnarray}
It is assumed that the scalar integral on the r.h.s. of (\ref{tensint})
has arbitrary masses and only after differentiation w.r.t. $m_i^2$ 
are these set to their concrete values.

To derive the explicit formula for the operator $T$, $L$ independent 
auxiliary vectors ${a_i}$ were introduced to represent the tensor 
in the integrand as  
\begin{eqnarray}
&&k_{1\mu_1} \ldots     k_{L\lambda_{n_L}}\!=\!\frac{1}{i^{n_1+...+n_L}}
\frac{\partial}{\partial { a_{1\mu_1}}}
  \ldots \frac{\partial}{\partial { a_{L\lambda_{n_L}}}}
\nonumber \\
&&~~\left.\times \!
 \exp \left[i ({ a_1}k_1+ \ldots+ { a_L} k_L)\right]
\right|_{  a_i=0 },
\nonumber 
\end{eqnarray}
and  after that the integral was transformed into an $\alpha$- parametric
representation.
From the parametric representation it is easy to deduce that
\begin{eqnarray}
T_{\mu_1 \ldots \lambda_{n_L}}(\{p_i\},\{\partial_j \},{ \bf d^+})
=\frac{e^{-iQ(\{ \{p\}, \alpha, \{0\} ) \rho} }{i^{n_1+ \ldots +n_N}}
 \nonumber \\ 
 \times  \frac{\partial}{ \partial { a_{1\mu_1}}} 
 \ldots  \left. \frac{\partial
~{ e}^{i[ Q(\{ p\},\alpha,\{ { a} \})] \rho }
}{ \partial { a_{L\lambda_{n_L}}}}
\right|_{ { a_j}=0~~\atop {\alpha_j=i \partial_j \atop
  \rho=i^L{\bf  d^+}} },
\nonumber 
\label{Ttensor}
\end{eqnarray}
where  $Q$ is a polynomial in $\alpha,~{ a}$ and $p_ip_j$. Applying  
$T$ gives a sum made of external momenta, $g_{\mu \nu}$ 
multiplied by scalar integrals with shifted $ d$. Advantages of the 
proposed tensor reduction are:
\begin{itemize}
\item no contractions with external momenta and metric tensor and
no solution of a linear system of equations are needed 
\item it is easy to select a particular tensor structure
\item
 representation in terms of integrals with shifted $ d$ is compact
\end{itemize}
Thus 
the problem is reduced to the evaluation of scalar integrals with 
different powers of scalar propagators 
and shifts in $ d$. They can be evaluated by 
applying generalized recurrence relations \cite{ovt1}.
To obtain these relations we start with the identity:
$$
\prod_{i=1}^{L} { \int} d^dk_i
\frac{\partial }{\partial k_{r\mu}}\!
\left\{ R_{\{ \mu\} }\left( \{ k \},\{ p \} \right)
\prod^{N}_{j=1}\!
P^{\nu_j}_{\overline{k}_j,m_j}\! \right\}\! \equiv  0,
$$
where $R$  is an arbitrary tensor polynomial. Such kind of identities
for finding relations between Feynman integrals was used in 
\cite{Petersson},\cite{tHV}. A systematic method was proposed  
in \cite{ibpm}. Upon differentiating two representations for scalar 
products can be used:
\begin{eqnarray}
&& {\rm a})~~ { integration~~ by ~~parts ~~method~ \cite{ibpm}.} 
   \nonumber \\
&&{ k_ip_l}=\frac12 (k_i^2+p_l^2-(k_i-p_l)^2),    \nonumber \\
&& {\rm b} )~~{generalized ~~recurrence ~~relations~ \cite{ovt1}}:
 \nonumber \\
&& {p_{l\mu}} \int{ k_{i \mu}} \prod^L_{r=1} d^dk_r  \prod^{N}_{j=1}
P^{\nu_j}_{\overline{k}_j,m_j} =  \nonumber \\
&&p_{l \mu} T_{\mu}( \{ p \}, \{ \partial \}, {\bf d^+} )
 \int \prod^L_{r=1}  d^dk_r \prod^{N}_{j=1} P^{\nu_j}_{\overline{k}_j,m_j}.
\nonumber 
\end{eqnarray}
Combining  these methods gives many recurrence relations
connecting integrals with different indices $\nu_j$  and shifts 
of  $ d $. 

\subsection{One-loop $n$-point integrals}
The method for the reduction of   one-loop tensor integrals 
$$
I_{n,r}^{(d)}=
\int \frac{d^d q}{i\pi^{{d}/{2}}} 
\prod_{j=1}^{n}\frac{q_{\mu_1} \ldots q_{\mu_r}}{c_j^{\nu_j}},
$$
with
$
c_j=(q-p_j)^2-m_j^2+i\epsilon,
$
was presented in \cite{fjt}. The integral $I^{(d)}_{n,r}$ can be written as
a combination of scalar integrals by using the relation
$$
I_{n,r}^{(d)}=T_{\mu_1 \ldots \mu_r} (\{ p_s \},\{ \partial_j \}, {\bf d^+})
~I_{n}^{(d)},
$$
where
\begin{equation}
I^{(d)}_n = \int \frac{d^d q}{i \pi^{{d}/{2}}} 
\prod_{j=1}^{n}\frac{1}{c_j^{\nu_j}},
\end{equation}
and
\begin{eqnarray*}
&&T_{\mu_1 \ldots \mu_r}(\{ p_j \},\{ \partial_j \}, {\bf d^+})
=\frac{1}{i^r} \\
&&\times \prod_{j=1}^r \frac{\partial}{\partial a_{\mu_j}}
\! \exp \left[i \left( \sum_{k=1}^{n}\!(ap_k) \alpha_k\!
 -\!\frac{a^2}{4} \! \right)
 \!\rho
    \right]\! 
 \left|_{ a_j=0 \atop {\alpha_j=i \partial_j \atop
  \rho=i {\bf d^+}} } \right. .
\end{eqnarray*}

Two recurrence relations are needed to reduce any scalar integral
$I_n^{(d)}$ to basic ones:
\begin{eqnarray}
&&2 \Delta_n \nu_j { \bf j^+} I_n^{({ d})}= 
 \sum^{n}_{k=1} (1+\delta_{jk}) \nonumber \\
&&  \times  \left( \frac{\partial \Delta_n}
 {\partial Y_{jk}} \right)
\left[ d - \sum_{i=1}^{n} \nu_i( {\bf k^-} { \bf i^+}+1)
             \right] I_n^{({ d})}, \\
&& (d\!-\!\!\sum_{i=1}^{n}\nu_i\!+\!1)I^{({ d+2})}_n= \nonumber \\
&&~~~~~~~~~~~~~~  \left[\frac{2 \Delta_n}{G_{n-1}}
\!+\!\!\sum_{k=1}^n \frac{(\partial_k \Delta_n)}{G_{n-1}}
{ \bf k^-}\!
  \right]I^{({ d})}_n,
\label{tworel}
\end{eqnarray}
where  $ \bf j^{\pm }$ etc. shift the indices $\nu_j \to \nu_{j } \pm 1$
and
\vspace{0.1cm}
\begin{eqnarray}
&&\Delta_n \equiv \Delta_n(\{p_1,m_1\},\ldots \{p_n,m_n\})= 
\nonumber \\
&&~~~~~~~~\left|
\begin{array}{cccc}
Y_{11}  & Y_{12}  &\ldots & Y_{1n} \\
\vdots  & \vdots  &\ddots & \vdots \\
Y_{1n}  & Y_{2n}  &\ldots & Y_{nn}
\end{array}
         \right|,~~~
\label{deltan}
\end{eqnarray}
\begin{eqnarray}
G_{n-1} \equiv G_{n-1}(p_1,\ldots ,p_n)= -2^n~~~~~~~~~~~
\nonumber \\
\times \left|
\begin{array}{ccc}
  \!p_{1n}p_{1n}  &\ldots & p_{1n}p_{n-1~n} \\
  \!p_{1n}p_{2n}  &\ldots & p_{2n}p_{n-1~n} \\
  \vdots  & \vdots & \vdots \\
  \!p_{1n}p_{n-1~n} &\ldots & p_{n-1~n}p_{n-1~n}
\end{array}
\right|, ~~
\label{Gn}
\end{eqnarray}
$$Y_{ij}=-p_{ij}^2+m_i^2+m_j^2,~~~p_{ij}=p_i-p_j.
$$ 
Here $p_i$ is the external momentum flowing through the $i$-th
line and $m_j$ is the mass of the propagator corresponding
to  the $j$-th line.

Relations similar to (\ref{tworel}) for the integrals with  
zero  Gram determinants were given in \cite{fjt}, \cite{BGH}.

In the next sections we will  also use an index notation for $\Delta_n$
and $G_n$
\begin{eqnarray}
&&\lambda_{i_1 i_2\ldots i_n}=\Delta_n(\{p_{i_1},m_{i_1}\},
\ldots ,\{p_{i_n},m_{i_n} \}),         \nonumber \\
&&g_{i_1  \ldots i_n}=G_{n-1}(p_{i_1},\ldots ,p_{i_n}).
\label{lage}
\end{eqnarray}
We shall use the index notation for
integrals obtained from $I_n^{(d)}$ by contracting some lines.
The characteristic variables occurring in our derivations will be  
ratios of Gram determinants and therefore it is worthwhile
to introduce the shorthand notation
\begin{equation}
r_{ij\ldots k}=-\frac{\lambda_{ij\ldots k}}{g_{ij\ldots k}}.
\end{equation}

\subsection{ Two-loop propagator integrals}
The reduction of the two-loop propagator type tensor integrals
to basic scalar integrals can be done by using the relation
\cite{ovt2}
\begin{eqnarray}
\int \! \int \!\frac{d^dk_1 d^dk_2}
  {c_1^{\nu_1}\!c_2^{\nu_2}\! c_3^{\nu_3}\!c_4^{\nu_4}\!c_5^{\nu_5}}
  ~k_{1\mu_1} \ldots k_{1\mu_r}
  k_{2\lambda_1}\! \ldots \!k_{2\lambda_s}
 \non \\
 =T_{\mu_1\ldots \lambda_s} (q,\{{\partial}_j\},
{\bf d^+}) \int \!\! \int \frac{d^dk_1 d^dk_2}
{c_1^{\nu_1} c_2^{\nu_2} c_3^{\nu_3} c_4^{\nu_4} c_5^{\nu_5}},
\label{twoloop}
\end{eqnarray}
\begin{eqnarray*}
c_1=k_1^2\!-\!m_1^2\!+\!i\epsilon,
~c_3=(k_1-q)^2\!-\!m_3^2\!+\!i\epsilon,\non \\
c_2=k_2^2\!-\!m_2^2\!+\!i\epsilon,~
c_4=(k_2-q)^2-m_4^2\!+\!i\epsilon, \non \\
c_5=(k_1-k_2)^2-m_5^2\!+\!i\epsilon. \non 
\end{eqnarray*}
An explicit form for the operator $T$ and all necessary recurrence 
relations are given in \cite{ovt2}. The recursive procedure allows one
to transform any diagram into a sum over 30 basic integrals:
$$
I^{(d)}(q^2)=\sum_{j=1}^{30} R_j(q^2,\{m_i^2 \},d) I^{(d)}_j(q^2).
$$
This  algorithm was implemented in \cite{SM} as a Mathematica package. 

The method of  tensor reduction \cite{ovt1} was used for the 
evaluation of radiative corrections to several important physical 
quantities. For example, for the evaluation of the 2-loop correction 
to the static potential in QCD \cite{YS} and the evaluation of the 2-loop 
correction in the gauge Higgs system in 3-dimensions \cite{FE}. Another
important application may be the evaluation of 2-loop correction to the  
Bhabha scattering \cite{SV}, \cite{Glover}.

\section{Explicit solution of $d-$ shift recurrence relations}

The solution of the $d-$ shift relations for one-loop propagator type 
integral was first discussed in \cite{ovt1}. Here we present the solution 
of the  relation (\ref{tworel}) for  the integral $I_n^{(d)}$  with 
arbitrary 
momenta and masses and first powers of scalar propagators. In this case 
\begin{equation}
  (d-n\!+\!1)I^{(d+2)}_n\!=\! \left[\!\frac{2 \Delta_n}
  {G_{n-1}}\!+\!\sum_{k=1}^n 
  \frac{\partial_k \Delta_n}{G_{n-1}}
  {\bf k^-}\!\right]I^{(d)}_n,
\label{reduceDtod}
\end{equation}
where the operator ${\bf k^-}$  removes the $k$-th line from
$I_n^{(d)}$.
If we assume that
$n-1$ point functions are already known then relation (\ref{reduceDtod}) 
is an inhomogeneous first order difference equation w.r.t. $d$. Methods of 
solution of this kind of equations are well described in
the mathematical 
literature \cite{Milne}. The  redefinition 
\begin{equation}
   I_{n}^{(d)}=\frac{1}{\Gamma\left(\frac{d-n+1}{2}\right)}
   \left( \frac{\Delta_n}{G_{n-1}}\right)^{\frac{d}{2}}
   \overline{I}_{n}^{(d)}
\label{defIbar}
\end{equation}
leads to a simpler equation  
\begin{eqnarray}
&&\overline{I}_{n}^{(d+2)}=\overline{I}_{n}^{(d)}
   +\frac{\Gamma \left(\frac{d-n+1}{2}\right)}
   {2 \Delta_n} \nonumber \\
&&~~~~\times
 \left(\frac{G_{n-1}}{\Delta_n}\right)^{\frac{d}{2}}
   \sum_{k=1}^{n}(\partial_k \Delta_n) {\bf k^{-}} I_{n}^{(d)}.
\label{Ibar}
\end{eqnarray}
Without loss of generality  $d$  can be parametrized as
$$
d=2l - 2 \varepsilon,
$$
with $l$ being an integer number and $\varepsilon$ an arbitrary small 
parameter. The solution of equation (\ref{Ibar}) then reads 
\begin{eqnarray}
  \overline{I}_{n}^{(2l-2\varepsilon)}\!\!=
  \tilde{b}_n(\varepsilon)+
  \sum_{r=0}^{l}
  \frac{\Gamma\left(r\!-\!1\!-\!\varepsilon\!-\!\frac{n-1}{2}\right)}
  {2\Delta_n} 
  \nonumber \\
\times  \left(\frac{G_{n-1}}{\Delta_n}\right)^{r-1-\varepsilon}
  \sum_{k=1}^{n}(\partial_k \Delta_n) {\bf k^{-}}I_{n}^{(2r-2
  -2\varepsilon)},
\label{solution1}
\end{eqnarray}
where $\tilde{b}_n(\varepsilon)$ is an $l$ independent constant.
By changing $\tilde{b}_n(\varepsilon)$ the solution (\ref{solution1}) 
can be rewritten in another form
\begin{eqnarray}
   \overline{I}_{n}^{(2l-2\varepsilon)} =
  \overline{b}_n(\varepsilon)
   -\sum_{r=0}^{\infty}
   \frac{\Gamma\left(r+\frac{d-n+1}{2} \right)}{2\Delta_n}
   \nonumber \\
\!\!\!\!\!\times
 \left(\frac{G_{n-1}}{\Delta_n}\right)^{r+\frac{d}{2}}
\sum_{k=1}^{n}(\partial_k \Delta_n) {\bf k^{-}}I_{n}^{(d+2r)}.
\label{solution2}
\end{eqnarray}
The  result for $I_n^{(d)}$ then reads
\begin{eqnarray}
I_n^{(d)}=\frac{1}{\Gamma \left(\!\frac{d-n+1}{2}\! \right)} 
\left(\! \frac{\Delta_n}{G_{n-1}}\!\right)^{\frac{d}{2}}
\!\overline{b}_n
  \!-\!\sum_{k=1}^{n}\! \left(\!\frac{\partial_k \Delta_n}
  {2 \Delta_n}\!\right) \!\nonumber \\
\times \sum_{r=0}^{\infty} \left(\!\frac{d\!-\!n\!+\!1}{2}
  \!\right)_r\left(\!\frac{G_{n-1}}{\Delta_n}\!\right)^{r}\!
  {\bf k^{-} } I^{(d+2r)}_n,
\label{npoint}
\end{eqnarray}
where  $\overline{b}_n$  can be determined from the asymptotic value 
of $I_n^{(d)}$ in the limit $d \rightarrow \infty$.

\subsection{Asymptotic form of $I_n^{(d)}$ at $d \rightarrow \infty$}
The asymptotic form of $I_{n}^{(d)}$ as $d \rightarrow \infty$
can be derived from its parametric integral representation.
$I_n^{(d)}$ can be written in a parametric form by employing, for
example, the following formula
\begin{eqnarray}
    \frac{1}{c_1 c_2 \ldots c_n}=\Gamma(n)
    \int_{0}^{1} \! \ldots \int_0^1 
    dx_1  \ldots dx_{n-1}~
\nonumber \\ 
\times \frac{x_1^{n-2} x_2^{n-3} \ldots x_{n-2}}
    {D^n},
\end{eqnarray}
with
\begin{eqnarray}
&&D=
[c_1\! x_1\!\ldots\! x_{n-1}\!\! \nonumber \\
&&+\!c_2x_1\!\ldots\! x_{n-2}
    (1\!-\!x_{n-1})\!+\!\ldots \!+\!c_n(1\!-\!x_1)],
\end{eqnarray}    
then shifting integration variable $q$ and  applying the formula
\begin{equation}
   \int \frac{d^dq}{[i \pi^{d/2}]}\frac{1}{{(q^2-m_i^2)}^{\alpha}}
   = \frac{(-1)^{\alpha}~\Gamma\left(\alpha-\frac{d}{2}\right)}
   {\Gamma(\alpha) (m_i^2)^{\alpha-\frac{d}{2}}}.
\label{tadpole}
\end{equation}
Any $n$-point function will be represented by a multiple
parametric  integral of the form:
\begin{eqnarray}
   I_n^{(d)}\!=\Gamma\left(n\!-\frac{d}{2}\right)
   \int_0^1 \!\!\ldots \!\int_0^1\! dx_1\ldots dx_{n-1} 
 \nonumber \\
 \times f(\{x\})
   (h_n(\{p_j,m_s\},\{x \}))^{\frac{d}{2}-n},
\label{parametric}
\end{eqnarray}
where $\{x\} = \{x_1,{\ldots} ,x_{n-1}\}$ and
$h_n$ is a polynomial.
In our opinion this particular parametric integral representation
is very convenient for performing asymptotic expansions.
The behavior of this  integral, as $d \rightarrow \infty$,
can be established by using asymptotic methods \cite{asybook}.
The main contribution to the integral will come from the maximum
of $h_n$. The maximum can be either on the boundary or in the interior 
of the integration  region. In case 
the solution $\{x \} = \{\overline{x}\}$ of the system of equations
\begin{equation}
\frac{\partial h_n}{\partial x_i}=0,  ~~~~~(1<i<n-1)
\end{equation}
lie in the interior of integration region then it is possible to show that
\begin{equation}
\overline{x}_i=\frac{\sum_{k=1}^{n-i} \partial_k \Delta_n}
{\sum_{j=1}^{n-i+1} \partial_j \Delta_n},
\label{extremum}
\end{equation}
and at this point
\begin{equation}
h_n(\{p_j,m_s\}, \{ \overline{x}\})= r_{1{\ldots} n}.
\end{equation}
In the domain of analyticity $h_n(\{p_j,m_s\},\{ x \})
 \geq 0$ and $I_n^{(d)}$ is an integral of  Laplace type.
In the domain of non-analyticity there are subdomains of the integration
region where $h_n(\{p_j,m_s\},\{ \overline{x} \}) < 0$  
and  so the integral $I_n^{(d)}$ has an imaginary part. For this
kinematic configuration $I_n^{(d)}$ may be represented as
\begin{eqnarray} 
&&I_n^{(d)}\!=\!\Gamma\left(\!n\!-\!\frac{d}{2}\!\right) 
\left \{\!
   \int \frac{ \{dx\}   \theta(h_n) f(\{x\})}
   {(h_n(\{p_j,m_s\},\{x\}))^{n-\frac{d}{2}}}\right.
   \nonumber \\
&& \!+\!\left(\cos \! \frac{\pi}{2}(d-2n) 
 + i \sin \frac{\pi}{2}(d-2n) \right)
 \nonumber \\
&& \times \!\left.
 \sum_j\! \int_{\Omega^-_j} \frac{ \{dx\} f(\{x\})}
   {\left| h_n(\{p_j,m_s\},\{x\})\right|^{n-\frac{d}{2}}}    
  \!\right \}.
\label{RePlusIm}
\end{eqnarray}
Here $\Omega^-_j$ are  subdomains of the integration region where $h_n<0$.
The boundaries of these subdomains can be found from the solutions
of  the equation
\begin{equation}
 h_n(\{p_j,m_s\},\{ \overline{x} \})=0.
\end{equation}
In (\ref{RePlusIm}) we  took the main branch of $h_n$ and  assumed that
the momenta and masses are real parameters. A concrete example of 
representation (\ref{RePlusIm}) will be given in the next section.

 The imaginary part of $I^{(d)}_n$ 
\begin{eqnarray} 
&& {\rm Im}    I_n^{(d)}=
 \frac{- \pi }{ \Gamma\left(\frac{d}{2}-n+1\right)}
\nonumber \\
&&~~\times \!\sum_j
 \int_{\Omega^-_j}\frac{ \{dx\} f(\{x \})}
   {\left| h_n(\{p_j,m_s\},\{x \} )\right|
        ^{n-\frac{d}{2}}}    
\label{ImPart}  
\end{eqnarray}
 is a sum of  integrals of Laplace type and therefore its
 asymptotic form at $d \rightarrow \infty$ can be established.
 
It should be noted that  representations like (\ref{RePlusIm}) 
are valid  for  multi loop integrals including  vacuum ones 
and also they can be used as an alternative to the method  for 
evaluating  imaginary part of the diagram with the help of
Cutkosky's rules.

The integrals (\ref{parametric}) are   multiple Laplace type integrals
of the form
\begin{equation}
    F(\lambda)=\int_{\Omega}f(x) \exp[\lambda S(x)]dx,
\end{equation}
where $\Omega$ is a bounded simply connected domain in $k$ dimensional
Euclidean space, $x=(x_1,\ldots x_k)$, $\lambda$ is a parameter,
$S(x)$ is a real function.
Supposing that the absolute maximum of $S(x)$ in $\Omega$ is achieved
only at the interior point $x=\overline{x}$, then at
$\lambda \rightarrow \infty$ 
\begin{equation}
   F(\lambda) \sim \exp[\lambda S(\overline{x})] 
   \sum_{r=0}^{\infty} a_k \lambda^{-r-\frac{k}{2}}.
\end{equation}
This expansion may be differentiated w.r.t. $\lambda$ any
number of times. The leading term of the expansion is
\begin{equation}
   F(\lambda)=\exp[\lambda S(\overline{x})]~ (2\pi/\lambda)^{\frac{k}
   {2}}~\frac{f(\overline{x})+O(\lambda^{-1})}{\sqrt{\left|{\rm det}
   S_{xx}(\overline{x})\right|}},
\label{maxinside}
\end{equation}
where $\left|{\rm det} S_{xx}(\overline{x})\right|$ is a Hessian
determinant.
The leading contribution from the integral at $d \rightarrow \infty$
comes from the region where $\left| h_n \right| $ has  maximal value.
If an extremum of $h_n$ occurs inside the integration region then we
must compare $|h_n|$ at the extremum point (\ref{extremum}) with 
its value on the boundary of the integration region and take the largest
one. If an extremum of $h_n$ lies outside the integration region then
the maximal value of $\left|h_n \right|$ will be reached on the boundary.
Since $h_n$ in the integral is multiplied by the factor
$ x_1^{n-2}x_2^{n-3}{\ldots} x_{n-2}, $ the non-zero contribution to 
the leading asymptotic  comes either from the boundary region 
$ x_{n-1}=0$ or from the boundary region where one of
$$
x_i=1, ~~(1 \leq i \leq n-1).
$$
At these $n$  boundary edges $h_n$ is equal to $h_{n-1}$ of
one of the $n$ integrals which one can obtain by contracting one of
the lines in $I_n$. Similar to $h_n$ the maximum value of 
these $h_{n-1}$ is equal either to their extremum value
\begin{equation}
h_{n-1}=\left| r_{i_1{\ldots} i_{n-1}}\right|
\end{equation}
or to the value on the boundary. Again at the boundary $h_{n-1}$
will be equal to the $h_{n-2}$  corresponding to integrals one
can obtain from $I_n^{(d)}$ by contracting two lines and so on. 
Splitting the integral for $h_n<0$ into  real and imaginary 
parts  (\ref{RePlusIm}) does not influence our considerations because 
splitting adds a boundary where $h_n$ vanishes  and therefore there 
will be no contribution to the asymptotic  value of the integral.

According to the previous discussion and from Eq.(\ref{npoint}) 
it follows  that in order to find  $\overline{b}_n$ the contribution
of $I_n^{(d)}$ at  $d \rightarrow \infty$ is needed only if it is 
proportional to  $r_{1{\ldots} n}^{d/2}$.  
Such an asymptotic value will occur  for 
kinematic configurations when all $\overline{x} \leq 1$ and $h_n$ 
has either maximum or negative minimum. In the latter case 
$I_n^{(d)}$ must be split into  real and imaginary parts according
to (\ref{RePlusIm}) and the asymptotic value of the term giving a
contribution 
proportional to $r_{1{\ldots} n}^{d/2}$ are to be determined.
It should be noted that the
contribution of order $r_{1{\ldots} n}^{d/2}$ may come from the analytic 
continuation of the sums in (\ref{npoint}). As will be seen in the next 
sections these sums  represent  hypergeometric functions.

From the formula for the  leading asymptotic of the multiple integral
(\ref{maxinside}) we obtain:
\begin{equation}
   \overline{b}_n=- (2\pi)^{\frac{n}{2}}
   ~\frac{\Gamma\left(1-\frac{d}{2}\right) 
   \Gamma \left(\frac{d}{2}\right)}{\sqrt{ \pi |G_{n-1}|} }
   ~\frac{i^d}
   {r_n^{\frac{n+1}{2}}},
\end{equation}
and therefore the general solution of  Eq.(\ref{reduceDtod})
for the kinematic configuration satisfying the conditions
$0 \leq \overline{x}_i\leq 1$
will be 
\vspace{1mm}
\begin{eqnarray}
 I_n^{(d)}\!=-(2 \pi)^{\frac{n}{2}}
  \frac{\Gamma\left(\!1\!-\!\frac{d}{2}\!\right) \Gamma \left(\!
  \frac{d}{2}
  \!\right)}{ \Gamma \left(\frac{d\!-\!n\!+\!1}{2}\right)}
  \frac{r_n^{\frac{d-n-1}{2}  }}
  {\sqrt{ \pi |G_{n-1}|}} \nonumber \\
 \nonumber \\
  \!-\!\sum_{k=1}^{n}\! \left(\!\frac{\partial_k \Delta_n}
  {2 \Delta_n}\!\right)\sum_{r=0}^{\infty} \left(\!\frac{d-n+1}{2}
  \!\right)_r  \nonumber \\
  \times \left(\!\frac{G_{n-1}}{\Delta_n}\!\right)^{r}
  {\bf k^{-} } I^{(d+2r)}_n,
\label{npointEucl}
\end{eqnarray}
where $(a)_r \equiv \Gamma(r+a)/\Gamma(a)$ is the Pochhammer symbol.
A representation of one-loop integrals in terms of 
Lauricella functions was given in \cite{adJMP}.


\section{2-point function}
Expression (\ref{npoint}) for $I_2^{(d)}$ includes two one-fold
infinite sums over tadpole integrals $I_1^{(d)}$ (see 
formula (\ref{tadpole})):
\begin{equation}
  I_1^{(d)}=-\Gamma\left(1-\frac{d}{2}\right)
  ~(m_i^2)^{\frac{d}{2}-1}.
\label{I1}
\end{equation}
It is convenient to label the lines of $I_2^{(d)}$ by $i,j$
because the integrals $I_2^{(d)}$   will be encountered in calculating 
$I_3^{(d)}$, {\ldots}   as a result of contraction of different 
lines. 

At $n=2$, substituting (\ref{I1}) into (\ref{npoint}) gives
\begin{eqnarray}
\frac{2 \lambda_{ij} I_2^{(d)} }{\Gamma\left(1\!-\!\frac{d}{2}\right)}
 \!=\!
 b_2\!+\! \frac{\partial_i \lambda_{ij}}{(m_j^2)^{1\!-\!\frac{d}{2}}}
   \sum_{r=0}^{\infty}\! \frac{\left(\! \frac{d-1}{2}
  \! \right)_r}{\left(\frac{d}{2}\right)_r}
   \left(\frac{m_j^2}{r_{ij}}\right)^r 
 \nonumber \\
 +\frac{\partial_j \lambda_{ij}}{(m_i^2)^{1-\frac{d}{2}}}
    \sum_{r=0}^{\infty}\! \frac{ \left(\frac{d-1}{2}\right)_r}
   {\left(\frac{d}{2}\right)_r}\left(\!\frac{m_i^2 }{r_{ij}}
   \!\right)^r. 
\label{i2sums}
\end{eqnarray}
One can easily recognize that each of the two infinite sums
can be written in terms of Gauss' hypergeometric function $_2F_1$. 
 The parametric integral  for $I_2^{(2)}$ is 
\begin{equation}
  I_2^{(d)}=\Gamma \left(2-\frac{d}{2}\right)\int_0^1 
  dx_1~h_2^{\frac{d}{2}-2},
\end{equation}
with
$$ h_2=p_{ij}^2x_1^2-x_1(p_{ij}^2-m_i^2+m_j^2)+m_j^2.$$
An extremum of $h_2$ is at
$   \overline{x}_1=
   (p_{ij}^2-m_i^2+m_j^2)/{(2p_{ij}^2)}.
$
The maximum of $h_2$ exists if 
\begin{equation}
  \frac{\partial^2 h_2}{\partial x_1^2}=2p_{ij}^2 < 0,
\end{equation}
i.e. only for Euclidean momentum.

The formula (\ref{ImPart}) gives the imaginary part of $I_2^{(d)}$ 
on the cut
\begin{equation}
{\rm Im}\!I_2^{(d)}\!=-\pi\!
\int^{x^+_1}_{x^-_1}\! dx_1
\frac{(\! (x_1\!-\!x^-_1)(x^+_1\!-\!x_1) )^{\frac{d}{2}-2}}
{p_{ij}^{4-d}~\Gamma \left( \frac{d-2}{2} \right)},
\end{equation}
where
$$
x^{\pm}_1=\frac{ p_{ij}^2-m_i^2+m_j^2 \pm \sqrt{-\lambda_{ij}}}
 {2p_{ij}^2}
$$
is the solution of the equation
\begin{equation}
h_2(\{p_{ij},m_i\},x_1)=0.
\end{equation}
The change of the integration variable $x_1=(x_1^+ -x_1^-) z +x_1^- $
makes the integration trivial and gives 
\begin{equation}
{\rm Im} I_2^{(d)} =\frac{-\pi~\Gamma \left( \frac{d-2}{2} \right)}
{p_{ij}^{4-d}~\Gamma \left( d-2 \right) }
\left( \frac{\sqrt{-\lambda_{ij}}}{p_{ij}^2} \right)^{d-3}.
\end{equation}
The real part of $I_2^{(d)}$ coming from the region where $h_2<0$
is
$$
{\rm Re} I_2^{(d)} = {\rm Im}I_2^{(d)} {\rm ctg} \frac{\pi}{2}(d-4).
$$
At large $d$ the asymptotic value of ${\rm Im}~I_2^{(d)}$  and
${\rm Re}~I_2^{(d)}$  can be easily found.
At $p_{ij}^2>(m_1+m_2)^2$ both $0 \leq x^{\pm}_1 \leq 1$.

Now several remarks concerning the evaluation of $b_2$ are in order.
From (\ref{npoint})  we see that at large
dimension $b_2 \sim  r_{ij}^{d/2}$.
Such a contribution may come either from the asymptotic value of
$I_2^{(d)}$ (this may happen only if $0< \overline{x}_1<1$)
at large $d$ or from the analytic  continuation of
infinite sums  when their expansion parameter 
exceeds 1.  To proceed further let us assume that $m_i^2>m_j^2$.
At $p_{ij}^2<m^2_j-m^2_i$ and $p_{ij}^2>(m_i+m_j)^2$
the value of $b_2$ is determined from the asymptotic  of
 $I_2^{(d)} \sim r_{ij}^{d/2}$. At $m^2_j-m^2_i<p_{ij}^2<(m_i-m_j)^2$
both $I_2^{(d)}$ and the infinite sums have no contributions
$\sim r_{ij}^{d/2}$ and therefore in this region $b_2=0$.
In the region $(m_i-m_j)^2 <p_{ij}^2<(m_1+m_2)^2$ the value
of $b_2$ is determined from the asymptotic value of infinite sums
at large $d$.

Combining  all values of $b_2$ and writing infinite sums
in terms of hypergeometric functions we obtained the following result
\begin{eqnarray}
&&\frac{2 \lambda_{ij}~ I_2^{(d)} }{\Gamma\left(1-\frac{d}{2}\right)} =
\nonumber \\
&& - \frac{\sqrt{\pi}\Gamma\left(\frac{d}{2}\right)}
 {\Gamma\left(\frac{d-1}{2}\right)}
 r_{ij}^{\frac{d-2}{2}}
 \left[ \frac{\partial_i  \lambda_{ij}}
        {\sqrt{1-\frac{m_j^2}{r_{ij}}}}
       +\frac{\partial_j \lambda_{ij}}
        {\sqrt{1-\frac{m_i^2}{r_{ij}}}} \right] \nonumber \\
&& \nonumber \\
&&~ + 
  \frac{\partial_i \lambda_{ij}}{(m_j^2)^{1-\frac{d}{2}}}
  \Fh21\Fdz{1,\frac{d-1}{2}}{\frac{d}{2}}
\nonumber \\
&&+ \frac{\partial_j \lambda_{ij}}
  {(m_i^2)^{1-\frac{d}{2}}}
  \Fh21\Fde{1,\frac{d-1}{2}}{\frac{d}{2}}.
\label{twopoint}
\end{eqnarray}
The analytic continuation of $_2F_1$ leads to
a compact form for $I_2^{(d)}$ (see also \cite{BDS})
\begin{eqnarray}
\frac{g_{ij}~I_2^{(d)}}
  { \Gamma\left(\!2\!-\!\frac{d}{2}\!\right)} = 
 \frac{\partial_i \lambda_{ij}}
     { (m_j^2)^{\frac{4-d}{2}}}\!
     \Fh21\Fdj{1,\frac{4-d}{2}}{\frac{3}{2}}
     \nonumber \\
  +\frac{\partial_j \lambda_{ij}}
      { (m_i^2)^{\frac{4-d}{2}}}\! 
     \Fh21\Fdi{1,\frac{4-d}{2}}{\frac{3}{2}}.
\end{eqnarray}
The expansion of $I_2^{(d)}$ to all orders in $\varepsilon=(4-d)/2$
is given in \cite{Andrey2000}.


\section{ 3-point function}


 The expression for $I_3^{(d)}$ includes summation over
$_2F_1$ coming from $I_2^{(d)}$. To derive an explicit formula we 
first  use the relation
\begin{equation}
(1-\!z)^{a}\!\Fh21\Ffz{a,b}{c}\!=\!\Fh21\Ffe{\!a,c\!-\!b}{c},
\end{equation}
in order to remove the factor $d$ from the second parameter of $_2F_1$.
Summing over  $I_2^{(d)}$ gives the Appell hypergeometric
functions \\
$F_3( (d-2)/2,1,1,1/2,d/2;x,y)$ defined as
\begin{eqnarray}
&&F_3(\alpha, \alpha', \beta,\beta',\gamma;x,y)=
    \nonumber \\
&& \sum_{m,n=0}^{\infty}\frac{(\alpha)_{m}(\alpha')_{n}
   (\beta)_{m}(\beta')_{n}}{(\gamma)_{m+n}~ m! ~n!}x^m y^n,
\end{eqnarray}
which may be reduced to $F_1$ by means of \cite{Erdeli}
\begin{eqnarray}
 && F_3(\alpha, \alpha', \beta,\beta',\alpha+\alpha';x,y)=
  \nonumber \\
&&  (1-y)^{-\beta'} F_1 \left(\alpha,\beta,\beta',\alpha+\alpha';
  x,\frac{y}{y-1}\right).
\end{eqnarray}
The Appell hypergeometric function $\!F_1\!$ defined as
\begin{eqnarray}
&&  F_1(\alpha,\beta,\beta',\gamma;x,y) = \nonumber \\
&&  \sum_{m,n=0}^{\infty}
 \frac{(\alpha)_{m+n}(\beta)_{m}(\beta')_{n}}
  {(\gamma)_{m+n} m! n!}x^m y^n.
\end{eqnarray}
Similar to $I^{(d)}_2$ we will label the lines of 
$I^{(d)}_3$ by $i,j,k$.
The result for $I_3^{(d)}$ valid  when
$h_3$ has maximum inside the integration region reads
%
%
\begin{eqnarray}
&&  \frac{\lambda_{ijk}}{\Gamma \left( 2-\frac{d}{2} \right)}
   I_3^{(d)}=2^{\frac32} \pi~\sqrt{-g_{ijk}}~ r_{ijk}^{\frac{d-2}{2}}
\nonumber \\
&&+\theta_{ijk}~\partial_k \lambda_{ijk}
  +\theta_{kij}~\partial_j \lambda_{ijk}
  +\theta_{jki}~\partial_i \lambda_{ijk},
\end{eqnarray}
where
\begin{eqnarray}
&&\lambda_{ij} \theta_{ijk}\!=\! \nonumber \\
&&\nonumber \\
&&\frac{(m_i^2)^{\frac{d-2}{2}}}
  {2(d-2)}
  \frac{\partial_j\lambda_{ij}}{ \sqrt{1-\frac{m_i^2}{r_{ij}}}}
  F_1\left(\!\frac{d\!-\!2}{2},1,\frac12,
  \frac{d}{2}; \frac{m_i^2 }{r_{ijk}},
  \frac{m_i^2}{r_{ij}}\!\right) 
\nonumber \\
&&+\frac{(m_j^2)^{\frac{d-2}{2}}}
  {2(d-2)}
  \frac{\partial_i\lambda_{ij}}{ \sqrt{1-\frac{m_j^2}{r_{ij}}}}
  F_1\left(\frac{d\!-\!2}{2},1,\frac12,
  \frac{d}{2}; \frac{m_j^2 }{r_{ijk}},
  \frac{m_j^2}{r_{ij}}\right) \nonumber \\
&&- \left[ \frac{\partial_i  \lambda_{ij}}
        {\sqrt{1\!-\!\frac{m_j^2}{r_{ij}}}}
       +\frac{\partial_j \lambda_{ij}}
        {\sqrt{1\!-\!\frac{m_i^2}{r_{ij}}}} \right]
       r_{ij}^{\frac{d-2}{2}}~
  \frac{\sqrt{\pi}~ \Gamma\left(\!\frac{d-2}{2}\!\right)}
  {4 \Gamma\left(\frac{d-1}{2}\right)} \nonumber \\
&&\times  
 \Fh21\Fdf{1,\frac{d-2}{2}}{\frac{d-1}{2}}.
\end{eqnarray}

The function $F_1$ has a simple integral representation
\begin{eqnarray}
F_1 \left( \frac{d-2}{2},1,\frac12,\frac{d}{2}; x,y \right)=
 \frac{d-2}{2} \nonumber \\
\times
\int_0^1 \frac{u^{\frac{d-4}{2}}} 
{(1-xu) \sqrt{1-yu} }du.
\label{intrepF1}
\end{eqnarray}

The polynomial 
\begin{eqnarray}
h_3=
  \!-x_1 x_2 (1-\!x_1) p^2_{13}-\!x_1^2x_2 (1-\!x_2)  p^2_{12} 
  \nonumber \\
    x_1 x_2 m_1^2-x_1 (1-\!x_1)(1-\!x_2)p^2_{23}\!
  \nonumber \\  
+x_1 (1-x_2) m_2^2+(1-x_1) m_3^2.
\end{eqnarray}
has a maximum at
$$
x_1=\frac{\partial_1 \Delta_3+\partial_2 \Delta_3}
{-G_2},
~~x_2=\frac{\partial_1 \Delta_3}
{\partial_1 \Delta_3+\partial_2 \Delta_3}
$$
if
\begin{eqnarray}
\label{h22}
&&\frac{\partial^2 h_3}{\partial x_2 \partial x_2}
 =2x_1^2 p_{12}^2 <0,\\
&& \nonumber \\
\label{hess}
&&\frac{ \partial^2 h_3}{\partial x_1 \partial x_1}
\frac{ \partial^2 h_3}{\partial x_2 \partial x_2}
-\left(\frac{ \partial^2 h_3}{\partial x_1 \partial x_2}\right)^2
 \nonumber \\
&&=-\frac{(\partial_1 \Delta_3+ \partial_2 \Delta_3)^2}{2G_2}>0.
\end{eqnarray}          
From (\ref{h22}),(\ref{hess}) it follows  that the  maximum will be 
achieved if
\begin{equation}
p_{12}^2<0,~~~~~~{\rm and}~~~~~~~G_2<0.
\end{equation}
The behavior of the integrand for $I_3^{(d)}$ having the maximum
inside the integration region is demonstrated in Fig.1.
\begin{center}
\vspace*{-2cm}
\vbox{
 \raisebox{9.0cm}{\makebox[0pt]{\hspace*{-3cm}$$}}
 \epsfysize=80mm \epsfbox{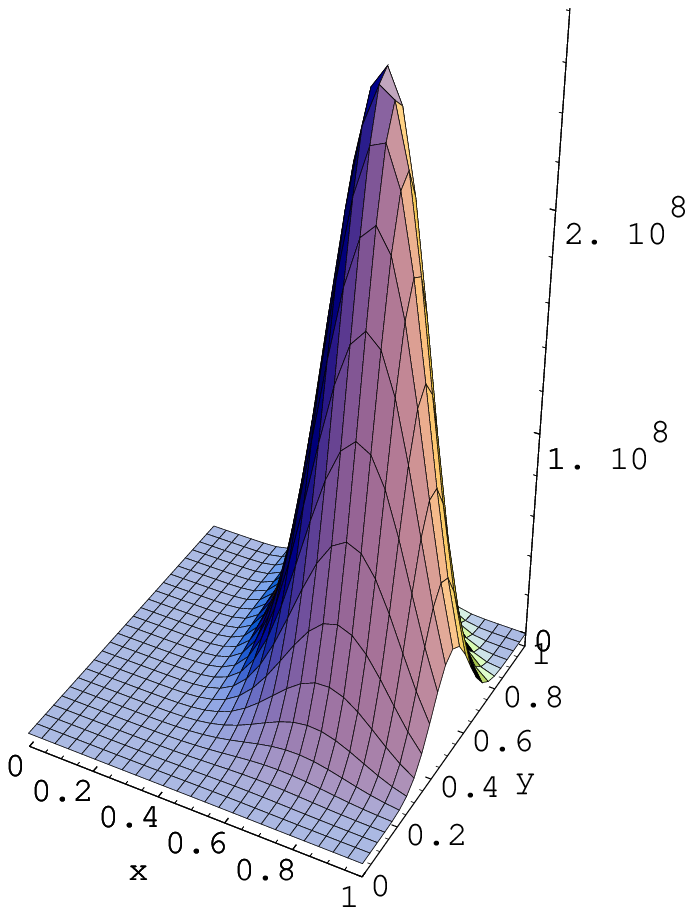}
     }
\end{center}
\noindent
\begin{center}
Fig. 1: Function $x_1h_3^{d/2-3}$ at $p_{13}^2=-10,~p_{23}^2=-5,~
p_{12}^2=-9$, $m_i^2=1$, $d=36$.
\end{center}

To evaluate $I_3^{(d)}$ in different kinematic regions an
analytic continuation of $_2F_1$ and $F_1$ is needed. The
corresponding formulae are given in \cite{fjt00}. The
representations of $I_3^{(d)}$ in terms of other hypergeometric
functions  can be found in \cite{BD}--\cite{AGO}.

\subsection{ 4-point function}

 The most problematic part  of the computation of  $I_4^{(d)}$
is the summation of terms from $I_3^{(d)}$ with $F_1$ function. 
To simplify the summation  we remove the factor $d$  in the
first argument of $F_1$ by applying the formula
\cite{Erdeli}
\begin{eqnarray}
 (1-x)^{\beta}(1-y)^{\beta'}F_1(\alpha,\beta, \beta',\gamma;x,y)
 = \nonumber \\
F_1\left(\gamma-\alpha, 
\beta,\beta',\gamma; \frac{x}{x-1},\frac{y}{y-1}\right).
\end{eqnarray}

The result for the $4$-point function then reads
\begin{eqnarray}
  \frac{\lambda_{ijkl}}{\Gamma\left( 2-\frac{d}{2}\right)}
  I_4^{(d)} =
      - 4\pi^{\frac32}~\sqrt{g_{ijkl}}~r_{ijkl}^{\frac{d-3}{2}}
      ~\frac{\Gamma\left(\frac{d-2}{2}\right)}
      {\Gamma\left(\frac{d-3}{2}\right)}
\nonumber \\
 \nonumber \\
+\phi_{ijkl} ~\partial_l \lambda_{ijkl}
     +\phi_{lijk} ~\partial_k \lambda_{ijkl}
     \nonumber \\
     +\phi_{klij} ~\partial_j \lambda_{ijkl}
     +\phi_{jkli} ~\partial_i \lambda_{ijkl},
\end{eqnarray}     
where
\begin{eqnarray}
&&\phi_{ijkl}=-\frac{\pi~\sqrt{2}}{\sqrt{g_{ijk}}}
  ~r_{ijkl}^{\frac{d-4}{2}}~
  \Fh21\Fbf{1,\frac{d-3}{2}}{\frac{d-2}{2}} \nonumber \\
&& \nonumber \\
&& ~~+~{\rm terms }~~{\rm with} ~~F_1~~{\rm and}~~ F_S
\end{eqnarray}
and $F_S$ is the Lauricella-Saran function \cite{Saran54} of three 
variables. In our case it has a simple integral representation:
\begin{eqnarray}
F_S=\frac{\Gamma \left(\frac{d}{2}\right)~(y-z)^{-\frac12}  }
{\Gamma\left(\frac{d-3}{2}\right) \Gamma\left(\frac32\right)}
\nonumber \\
\times \int_0^1 \frac{\arcsin \sqrt{\frac{(y-z)t}{1-tz}}
 (1-t)^{\frac{d-5}{2}}}{(1-x+tx) \sqrt{1-ty}} ~dt .
\end{eqnarray}
All other details one can find in \cite{fjt00}. 
\smallskip
\section{Conclusions}
The recurrence relations with respect to the space-time dimension
turn out to be useful for the reduction of tensor integrals.
The relations for master integrals in different dimensions
are rather simple even for integrals with several
momenta and masses. Preliminary investigation of
recurrence relations w.r.t.  $d$ for the two-loop propagator
type integrals revealed that scalar integral in (\ref{twoloop})
with all $\nu_j=1$ can be written as multiple hypergeometric 
series.
Asymptotic expansion  at large $d$ may be used as yet
another tool for approximate evaluation of Feynman integrals.
As is known such an expansion works well in quantum mechanics and
in quantum field theory in lattice calculations.

{\bf Acknowledgements}. I wish  to thank organizers of the
conference ``Loops and Legs 2000'' Johannes Bl\"umlein and Tord
Riemann for the useful and well organized conference.
I am thankful to C. Ford and  V. Ravindran for carefully reading
 the manuscript and useful remarks.
I would like also to thank the DFG for  financial support.

\end{document}